\documentclass[aps,pre,twocolumn,groupedaddress,showpacs,floatfix]{revtex4}
\usepackage{amsmath}
\usepackage{amssymb}
\usepackage{graphicx}

\begin{document}

\title{The Hierarchical Backbone of Complex Networks} 

\author{Luciano da Fontoura Costa} 
\affiliation{Institute of Physics of S\~ao Carlos. 
University of S\~ ao Paulo, S\~{a}o Carlos,
SP, PO Box 369, 13560-970, 
phone +55 162 73 9858,FAX +55 162 71
3616, Brazil, luciano@if.sc.usp.br}

\date{\today}

\begin{abstract}   

Given any complex directed network, a set of acyclic subgraphs --- the
hierarchical backbone of the network --- can be extracted that will
provide valuable information about its hierarchical structure.  The
current paper presents how the interpretation of the network weight
matrix as a transition matrix allows the hierarchical backbone to be
identified and characterized in terms of the concepts of
\emph{hierarchical degree}, which expresses the total number of
virtual edges established along successive transitions, and of
\emph{hierarchical successors}, namely the number of nodes accessible
from a specific node while moving successive hierarchical levels.  The
potential of the proposed approach is illustrated with respect to word
associations and gene sequencing data.

\end{abstract}


\maketitle

Although the study and characterization of complex networks
(e.g. \cite{Albert_Barab:2002, Newman:2003}) has often relied on
simple measurements such as the average node degree, clustering
coefficient and average length, such features do not provide direct
insights about several relevant properties of the analyzed networks.
While such limitations have been acknowledged from time to time and
complementary measures have been duly proposed in the literature,
including the connectivity correlation \cite{Satorras:2002} and
betweeness centrality \cite{Goh_etal:2001}, relatively lesser
attention has been given to measurements or algorithms capable of
comprehensively expressing the hierarchical structure of complex
networks, and only more recently attention has been focused on their
hierarchy \cite{Ravasz_Barab:2002, ravasz_etal:2003, Caldarelli:2002,
Barab_Oltvai:2003, Trusina_etal:2003, Boss_etal:2003,
Barthelemy_etal:2003, Zhou:2003, Vazquez:2003, Steyvers:2003}.
Indeed, even if such networks often involve cycles, their hierarchical
structure can be identified and characterized in terms of concepts
such as the \emph{hierarchical successors} and \emph{hierarchical
degrees}, herein introduced.  The present work concentrates on
directed, weighted complex networks (digraphs), illustrating the
potential of the suggested concepts and algorithms with respect to
complex networks derived from word association psychophysical
experiments and gene sequencing in zebrafish.  It is argued that the
hierarchical degree density represents a natural extension of the
classical node degree density, being capable of providing additional
information about the network hierarchy and connectivity.

Let the complex weighted directed network (or digraph) $\Gamma$ be
represented in terms of its nodes $k=1, \ldots, N$, and the directed
edges expressed as pairs $(i,j)$, with respective weights $w_{i,j}$,
which can be represented into the weight matrix as $W(j,i)=w_{i,j}$.
Self-connections are not considered in this work.  Let $\delta_T(a)$
be the operator acting elementwise over its argument $a$ --- which can
be a scalar, vector or matrix --- in such a way that to each resulting
element is assigned the value one whenever the respective element of
$a$ has absolute value larger than the specified threshold $T$; for
instance, $\delta_2(\vec{x})=(4 ,-1, 0, -3, 2)=(1,0,0,1,0)$.  Thus,
the adjacency matrix underlying the complex network $\Gamma$ can be
expressed as $\breve{W}=\delta_0(W)$.  The \emph{indegree}
$Id_k=\sum_{i=1}^{N}W(k,i)$ of node $k$ in such a complex network
corresponds to the total weight of incoming edges, and the
\emph{outdegree} $Od_k=\sum_{i=1}^{N}W(i,k)$ to the total weight of
outgoing edges.  If the operator $\Omega_A^t$ acts over a matrix $A$
as given by Equation~\ref{eq:omega}, the hierarchical successors of
node $k$ reached along $t$ transitions from the initial state
correspond to the non-zero entries of the vector $\vec{s}(t)$
calculated as in Equation~\ref{eq:succ_aft}, where $\vec{x}_k^{(0)}$
has all elements zero except the $k-$element, which is set to $1$. The
successors of $k$ reached exactly at the transition $t$ can be
therefore determined by Equation~\ref{eq:succ_at}. The number of
successors, without repetitions, of node $k$ at $t$, hence $U_k(t)$,
provides valuable information about the ramifications of the network
along its hierarchical levels $t$.  For instance, if the network is a
complete branching tree with $r$ branches at each fork, the root node
$k$ leads to $U_k(t) = r^t$, i.e. a power-law.  Even for a network
containing cycles, the successors of each of its constituent nodes $k$
will have a finite maximum depth $D_k$ defined as the value of $t$
such that $\vec{u}_k(t+1)=\vec{0}$.  The nodes identified by non-zero
elements of $\vec{s}_k^{(D_k)}$, together with the edges between them,
define a subgraph of $\Gamma$ which is henceforth called the
\emph{k}-component $C_k$ of $\Gamma$, whose number of nodes is denoted
by $N_k$.

\begin{eqnarray}
  \Omega_A^t = \sum_{i=1}^{t} A^i  \label{eq:omega} \\
  \vec{s}_k^{(t)} = \delta( \Omega_W^t \vec{x}_k^{(0)}) \label{eq:succ_aft} \\
  \vec{u}_k^{(t)} = \vec{s}_k^{(t)} - \vec{s}_k^{(t-1)} \label{eq:succ_at}
\end{eqnarray}

Given any \emph{k}-component $C_k$ of $\Gamma$, it is possible to
obtain a respective acyclic graph $G_k$ \footnote{A network is said to
be acyclic iff it contains no cycles.}, henceforth called the
\emph{k-th hierarchical component} of $\Gamma$, by using the following
algorithm:

\begin{tabbing}
Inc\=lude all nodes reachable from $k$ after one transition \\
   \> into the set $S$, and store the respective outgoing edges  \\
   \> of $k$ into the new adjacency matrix $\breve{W}_k^G$; \\
Remove all incoming edges to $k$;  \\
Whi\=le $S$ is not empty \\
   \> Remove all edges between the elements in $S$;  \\
   \> For \=each element $p$ of $S$, in any order:  \\
   \>    \> Incl\=ude all nodes reachable from $p$ after one \\
   \>    \>     \> transition into the set $R$, incorporating \\
   \>    \>     \> the outgoing edges of $p$ into the new \\
   \>    \>     \> adjacency matrix $\breve{W}_k^G$; \\
   \>    \> Remove all incoming edges to $p$;  \\
   \> Do $S=R$;  \\
\end{tabbing}

\begin{figure}
 \begin{center} 
   \includegraphics[scale=.6,angle=-90]{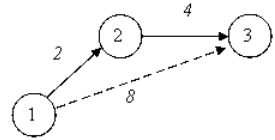}
   \caption{A simple graph illustrating the existence of a 
   virtual edge between nodes $1$ and $3$, defined at
   $t=2$ by the fact that $W^2(3,1)=8$.~\label{fig:virtual}} 
 \end{center}
\end{figure}

\begin{figure}
 \begin{center} 
   \includegraphics[scale=.6,angle=-90]{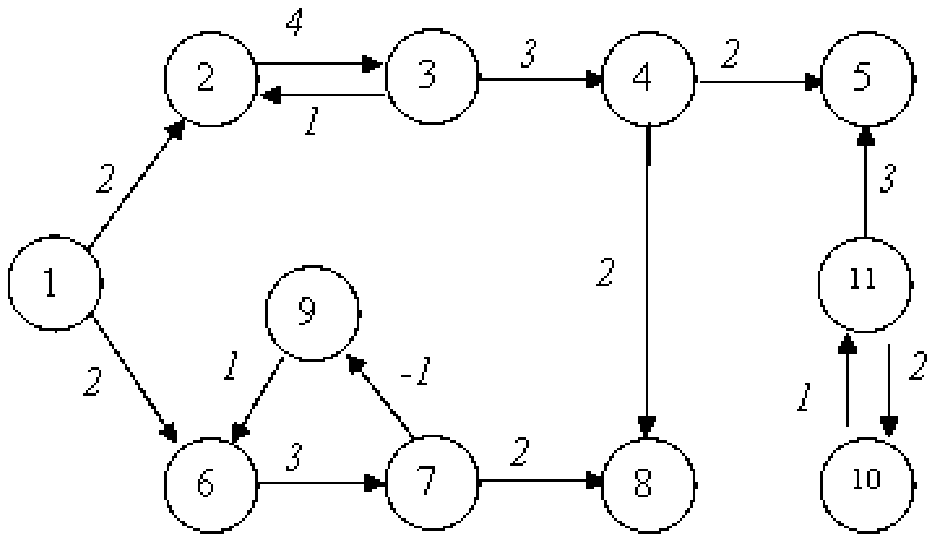}  \\ (a) \\
   \includegraphics[scale=.55,angle=-90]{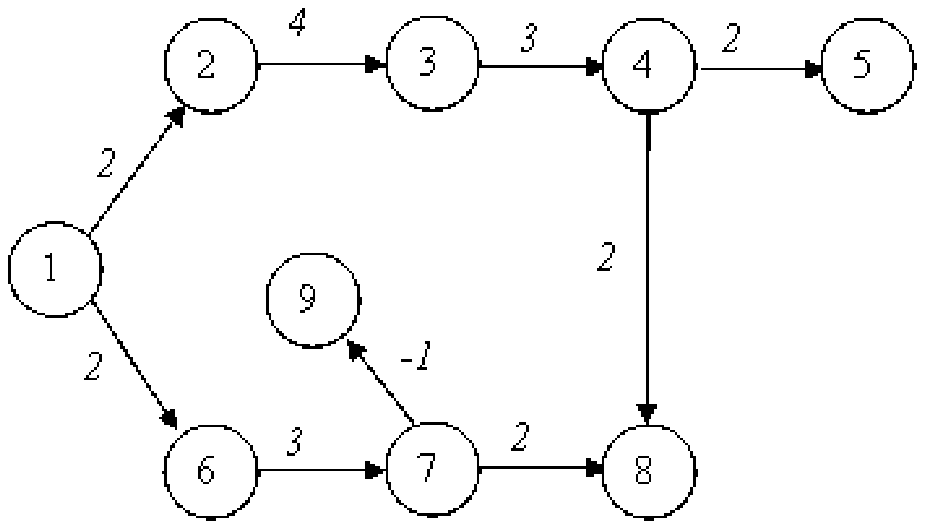}
   \includegraphics[scale=.55,angle=-90]{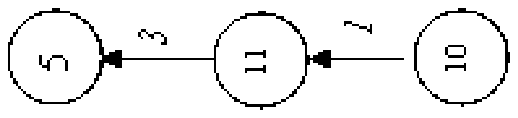} \\
  \end{center}
   \hspace{2cm} (b) \hspace{3cm} (c)
  \begin{center}
   \caption{A simple network (a) and its hierarchical components
      for $k=1$ (b) and $10$ (c). The dominant hierarchical component,
      containing 9 nodes, is shown in (b). We have $D_1=4$, 
      $D_{10}=2$, $N_1=8$, $N_{10}=2$, $\vec{u}_1^{(t)}=2,2,3,1$,
      $\vec{u}_{10}^{(t)}=1,1$ $h_1^{(t)}=4,14,30,48$ and
       $h_{10}^{(t)}=1,3$. ~\label{fig:ex}} 
   \end{center}
\end{figure}

The weight matrix of the acyclic graph $G_k$ is now given as $W_k^G =
\breve{W}_G^k .* W$, where `$.*$' is the elementwise product between
two matrices. The set of such acyclic subgraphs $G_k$, $k = 1, \ldots,
N$ is henceforth understood as the \emph{hierarchical backbone} of the
complex network $\Gamma$.  Since the network $\Gamma$ produces $N$
hierarchical components, some criterion can be used to identify those
most significant, such as the largest number of nodes.  It is now
possible to define the \emph{hierarchical outdegree} of each node $k$
of $G_k$ at the transition $t$, hence $h_k^{(t)}$, as the sum of the
weights of the virtual edges established between $k$ and any other
nodes of $G_k$ exactly at that instant.  As the existence of a
\emph{virtual link} from node $k$ to $j$ is understood to occur
whenever $W^t(j,k) \neq 0$, as illustrated in
Figure~\ref{fig:virtual}, the hierarchical degree of $k$ in $G_k$ can
therefore be obtained from Equation~\ref{eq:hd_at}.  The cumulative
hierarchical degree of node $k$ after \emph{t}-transitions, hence
$H_k^{(t)}$, is given by Equation~\ref{eq:hd_after}.  It is suggested
in this paper that the hierarchical degree, as well as its cumulative
version, provide a natural and powerful extension of the traditional
concept of node degree (which coincides with $h_k(t=1)$), capable of
providing more comprehensive information about the hierarchical and
connectivity properties of the graph topology.  Observe that the
number of nodes in the subgraph $G_k$ can be calculated as in
Equation~\ref{eq:Nc}.  Figure~\ref{fig:ex} shows a simple network (a)
and its hierarchical components $G_1$ and $G_{10}$.

\begin{eqnarray}
  h_k^{(t)} = \sum_{i=1}^N (W_k^G)^t(i,k) \label{eq:hd_at} \\ 
  H_k^{(t)} = \sum_{i=1}^N \Omega_{W_k^G}^t(i,k) 
           \label{eq:hd_after} \\ 
  N_k = 1+\sum_{i=1}^{N} \delta(\vec{s}_k^{(D_k)}(i)) \label{eq:Nc}
\end{eqnarray}

The hierarchical degree of a node for a specific $t$ can also be
understood as the traditional degree of that node in an
\emph{augmented network} including all virtual connections established
along the successive interactions.  The adjacency matrix of such an
enlarged network can be calculated as $A_k^G = (W_k^G)^t$.

In order to illustrate the potential of the above proposed concepts
and algorithms for hierarchical characterization of complex networks,
they have been applied to experimental data for word associations
\cite{Costa_what:2003} and zebrafish gene sequencing.  In the word
associations experiment, a graph was obtained for a human subject
through a psychophysical experiment as described in
\cite{Costa_what:2003}.  Starting with the word \emph{sun}, the
subject was requested to enter the first word that came to his mind
after reading the program-supplied word.  Except for the first word,
all other words presented by the program are drawn from those
previously supplied by the subject.  Statistical methods are applied
so as to ensure that every considered word is presented a similar
number of times.  A typical sequence obtained by this experiment is:
sun $\mapsto$ \emph{round}; round $\mapsto$ \emph{circle}; sun
$\mapsto$ \emph{gold}; gold $\mapsto$ \emph{yellow}; $\ldots$.

\begin{figure}
 \begin{center} 
   \includegraphics[scale=.6,angle=-90]{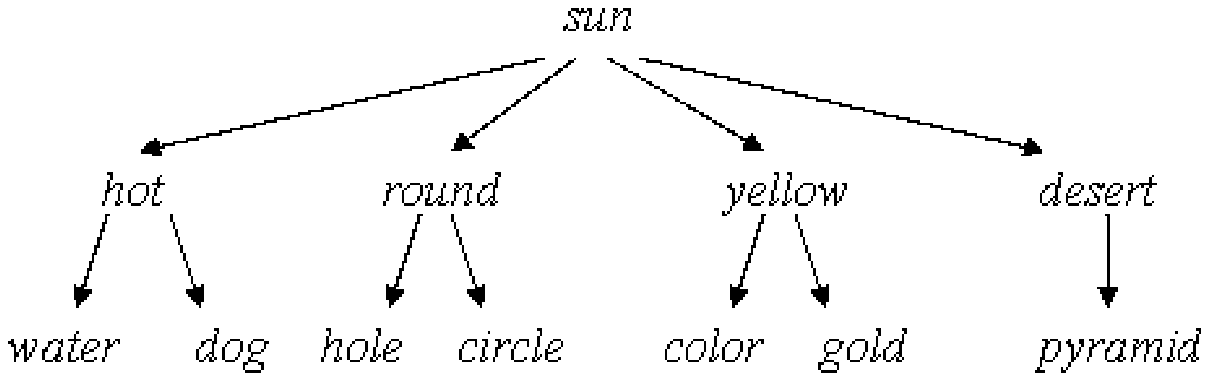} \\
   (a) \\
   \includegraphics[scale=.6,angle=-90]{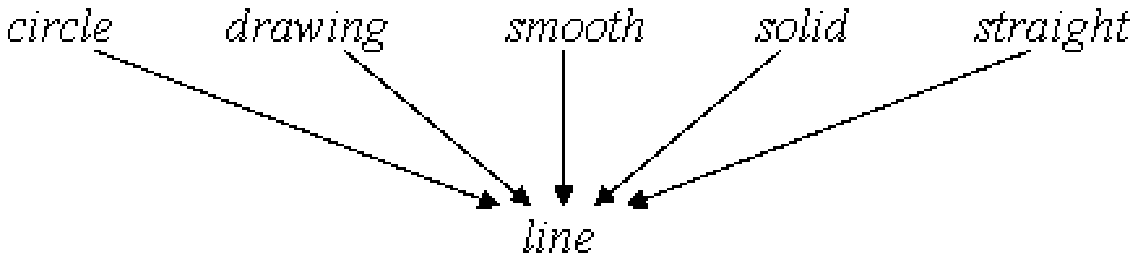} \\
   (b) 
   \caption{The first three levels of the hierarchy obtained
   for word associations starting at the node defining the largest 
   cluster size (a); and the last two hierarchical levels converging
   to the word \emph{line}, obtained from the inverted 
   associations.~\label{fig:word_hier}} \end{center}
\end{figure}

\begin{figure}
 \begin{center} 
   \includegraphics[scale=.6,angle=-90]{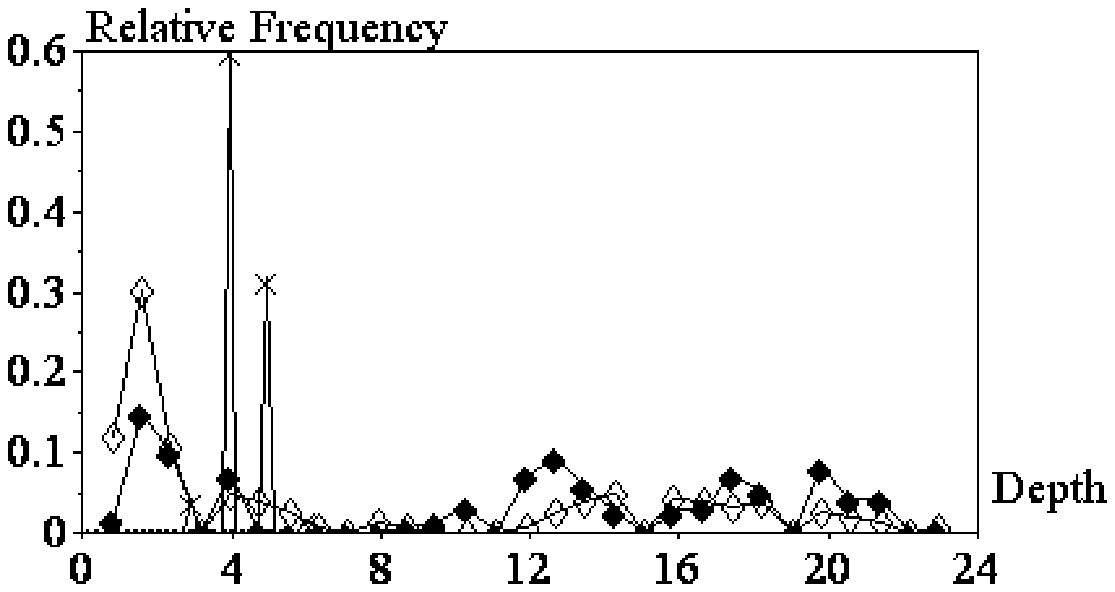} \\
   (a) \\ 
   \includegraphics[scale=.6,angle=-90]{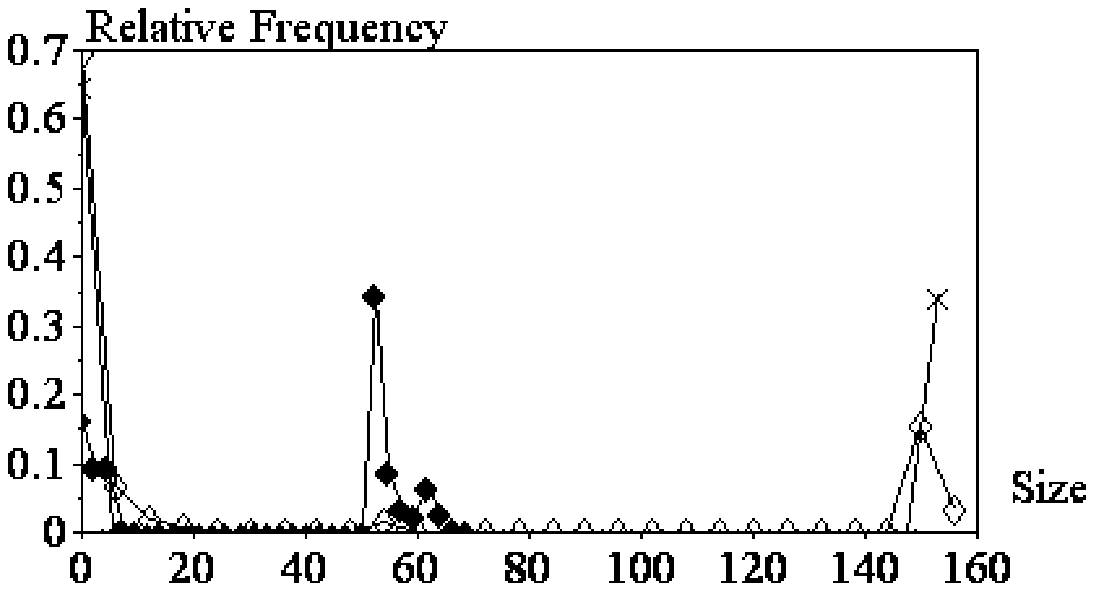} \\ 
   (b)
   \caption{Densities of: (a) depths $D_k$, and (b) 
   cluster sizes $N_k$, for the three considered cases: 
   filled diamond = word association; diamond = inverse word 
   association; and $\times$ = gene sequences.~\label{fig:resa}}
   \end{center}
\end{figure}

\begin{figure}
 \begin{center} 
   \includegraphics[scale=.5,angle=-90]{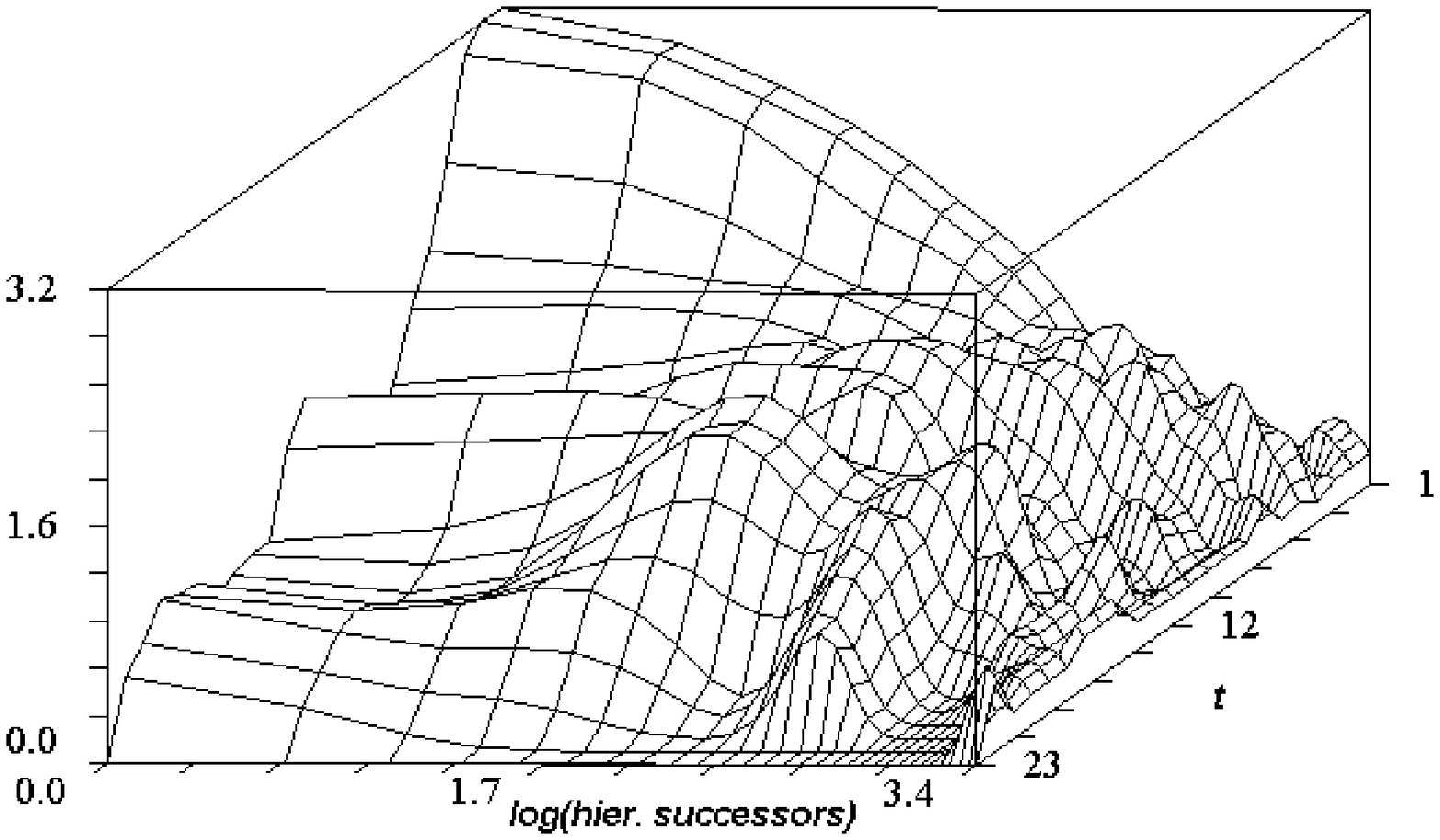} \\
   (a) \\
   \includegraphics[scale=.5,angle=-90]{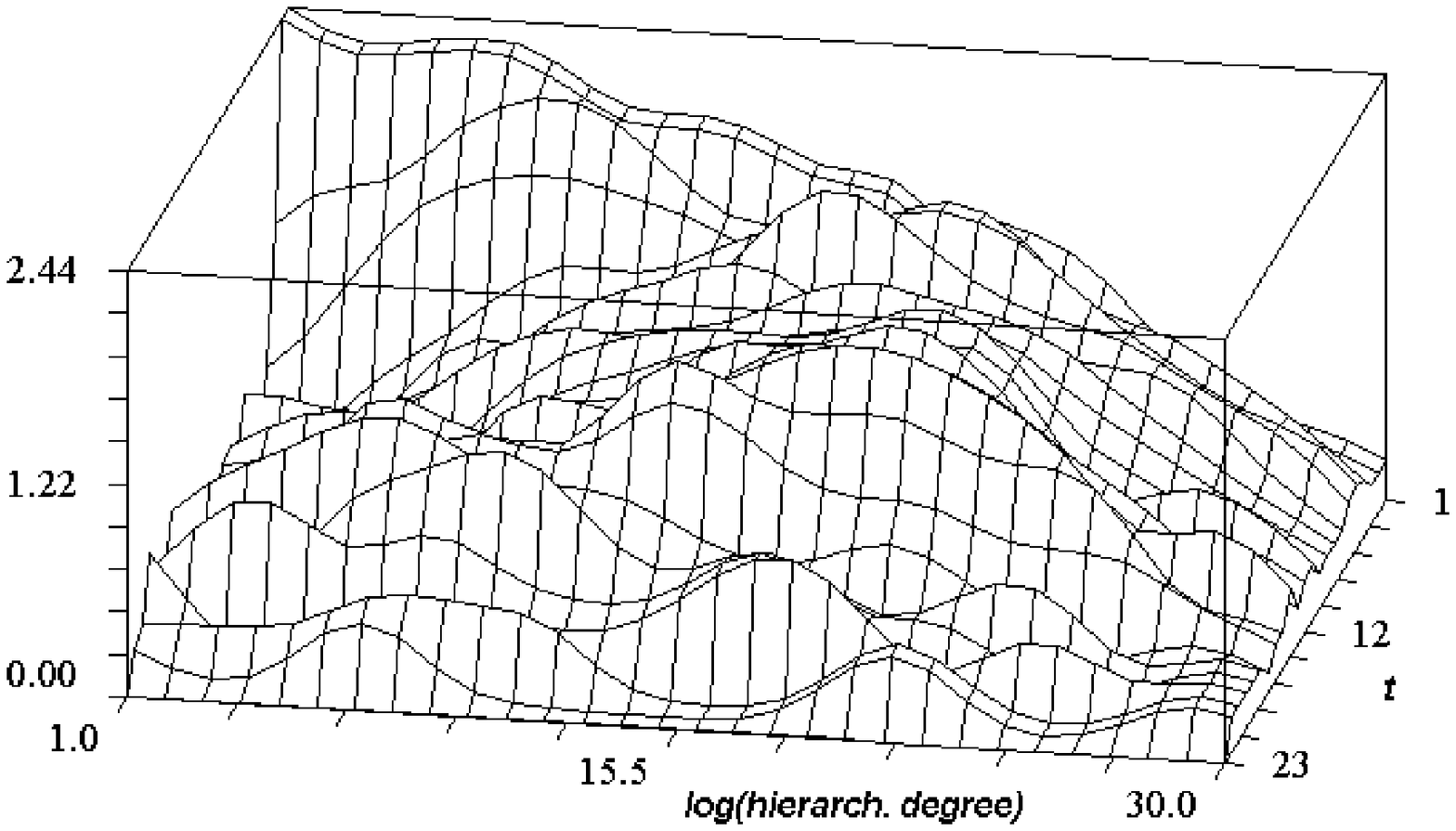} \\
   (b)
   \caption{Dilog densities for hierarchical
   successors (a) and hierarchical degrees (b) for the
   inverse word association data.~\label{fig:dens_word}} 
 \end{center}
\end{figure}

Once a large number of such associations are obtained (1930 in the
considered experiment), a weighted directed graph is determined by
representing each of the words (250 for this experiment) as nodes and
the number of specific associations between two words as an edge
between the respective nodes, weighted by the number of
associations. Provided the outgoing weights are normalized, this type
of graph can be understood as a Markov chain. It is suggested that
this graph provides interesting information about the tendencies of
the subject to associate words and concepts, paving the way to a
series of possibilities such as the identification, from the
respective network topological properties, of the author of the
associations.  As the node indegrees in such a graph were found
previously \cite{Costa_what:2003} to follow a power law, the
transposed weight matrix is also considered in this work.  In
addition, as the topology of a network can be severely modified by
addition of an incorrect edge (e.g. an eventual mistake while making
the associations), the graph obtained from that experiment was
thresholded such that only edges with weight larger than one were
considered, yielding the weight matrix $W$.  The maximum total cluster
size, i.e. $69$ was obtained for the initial word, i.e. \emph{sun},
whose derived hierarchical structure is shown in
Figure~\ref{fig:word_hier}(a), which is not the same as the word
\emph{sky} corresponding to the maximum traditional degree. Such
results reflect the fact that, by appearing soon at the beginning of
the experiment, this word implied a larger number of indirectly
related words along the rest of the experiment, even though the words
were presented with similar frequencies. A maximum cluster size of 154
was obtained for the inverted associations, corresponding to the word
\emph{line}, as illustrated in Figure~\ref{fig:word_hier}(b).
Interestingly, despite such a trend toward hierarchy, the obtained
graph also incorporate several cycles defined between different
hierarchical levels (see \cite{Costa_what:2003}), which were duly
eliminated in the present investigation.  The plots of relative
frequencies showing the hierarchical depths and the cluster sizes are
shown in Figure~\ref{fig:resa}.  The dilog plot of the densities
obtained for the number of hierarchical successors and hierarchical
degree considering all nodes in the backbone and $t < 24$ are shown in
Figure~\ref{fig:dens_word}(a-b).

The other experimental data considered in this paper regards zebrafish
(\emph{Brachydanio rerio}) gene sequencing data, obtained from the
\emph{NIH Zebrafish Gene Collection} repository \cite{zebrafish} (file
\verb+dr_mgc_cds_aa.fasta+).  The raw data consisted of sequences of
aminoacid for the first 892 genes in that file, which were organized
into subsequent pairs and the number of successive occurrences of such
pairs were calculated and used to build a complex weighted network,
which included 400 nodes ($20^2$ aminoacids) and 112582 sequential
associations.  The weight matrix was thresholded at 5000.  The
obtained depth and cluster size densities are shown in
Figure~\ref{fig:resa}(a) and (b).  The maximum cluster was obtained
for 147 nodes, which formed a densely connected cluster. The node
leading to the maximum cumulative hierarchical degree $H_k^{(4)}$
corresponded to the aminoacid pair \emph{QK}, i.e. glycine and lysine.

It is clear from the depth density plot in Figure~\ref{fig:resa}(a)
that the direct and inverted word associations led to similar profiles
characterized by wide distribution of depths, while the gene
sequencing data implied a small number of depth values, with a peak at
4.  This fact is explained by the similar weights for the aminoacid
pair associations, which produced an abrupt transition of the
connectivity of that graph as the threshold was raised.  The higher
density of shallow nodes (i.e. with depth between 0 and 4) observed
for the inverse word associations indicates the fact that shorter
streams of associations were established with the words entered by the
subject at the last stages of the experiment.  Interestingly, several
words were characterized by long streams of associations.  The size
density plot in Figure~\ref{fig:resa}(b) shows an interesting high
density of cluster sizes between 50 and 60.  This fact is possibly
explained by the fact that one of such clusters acts as an attractor
to several of the network nodes.  A large cluster was obtained for the
inverse word associations, which reflects the asymetric nature of the
underlying graph.  A large peak was observed for the gene sequences
which, considered jointly with Figure~\ref{fig:resa}(b), corroborates
the fact that most nodes in this case are accessible through a few
hierarchical levels.  It is observed from Figure~\ref{fig:dens_word}
(a) that the hierarchical degree tends to produce peaks at specific
hierarchical levels $t$, indicating the presence of bottlenecks along
the network.  While the number of small valued hierarchical successors
tend to decrease with $t$, the opposite behavior is observed for the
larger numbers.  The hierarchical degree shown in
Figure~\ref{fig:dens_word}(b) presents a rich structure indicating
that this measure tends to vary considerably with $t$.

In summary, it has been shown that the hierarchical underlying
structure of complex networks provide valuable information about its
topology that can not be fully appreciated by considering traditional
measurements.  The proposed decomposition of the network into maximal
hierarchical components, as well as the concepts of hierarchical
successors and hierarchical degree, allowed a more comprehensive
characterization of the hierarchical structure of complex networks.
It is expected that such tools will complement the characterization of
complex networks possibly incorporating strong hierarchical backbone,
such as the internet, metabolic networks, linguistic and social and
economical systems.

\begin{acknowledgments}
The author is grateful to FAPESP (proc. 99/12765-2) and CNPq
(proc. 301422/92-3) for financial support.
\end{acknowledgments}

 
\bibliography{hiermat}

\end{document}